\documentclass[fleqn,10pt]{wlscirep}
\usepackage[utf8]{inputenc}
\usepackage[T1]{fontenc}

\usepackage{graphicx} 

\usepackage{amsmath,amsthm,amssymb,mathrsfs}
\usepackage{bm}

\usepackage{color}

\title{Spintronic virtual neural network by a voltage controlled ferromagnet for associative memory}

\author[1*]{Tomohiro Taniguchi}
\author[2]{Yusuke Imai}
\affil[1]{National Institute of Advanced Industrial Science and Technology (AIST), Research Center for Emerging Computing Technologies, Tsukuba, Ibaraki 305-8568, Japan}
\affil[2]{Graduate School of Information Science and Technology, The University of Tokyo, Bunkyo-ku 113-8656, Tokyo, Japan}

\affil[*]{tomohiro-taniguchi@aist.go.jp}

\begin{abstract}
Recently, an associative memory operation by a virtual oscillator network, consisting of a single spintronic oscillator, was examined to solve issues in conventional, real oscillators-based neural networks such as inhomogeneities between the oscillators. 
However, the spintronic oscillator still carries issues dissipating large amount of energy because it is driven by electric current. 
Here, we propose to use a single ferromagnet manipulated by voltage-controlled magnetic anisotropy (VCMA) effect as a fundamental element in a virtual neural network, which will contribute to significantly reducing the Joule heating caused by electric current. 
Instead of the oscillation in oscillator networks, magnetization relaxation dynamics were used for the associative memory operation. 
The associative memory operation for alphabet patterns is successfully demonstrated by giving correspondences between the colors in a pattern recognition task and the sign of a perpendicular magnetic anisotropy coefficient, which could be either positive or negative via the VCMA effect. 
\end{abstract}

\begin{document}

\flushbottom
\maketitle
%
%


Emulating associative memory operation in the human brain by electrical devices has been investigated since the 1970s \cite{nakano72,kohonen72,anderson72}. 
Several models inspired by neural and/or synaptic activities, such as the Hopfield model \cite{hopfield82}, and their experimental implementations have been developed \cite{amari77,hemmen86,mceliece87,waugh90,morita93,yoshizawa93,bolle93,krebs99,mcgraw03,zhao04}. 
For example, the associative memory operation was recently achieved by using nanometer-scale ferromagnetic memory as synapses \cite{borders17}. 
The associative memory operation has also been examined by another model, called coupled oscillator networks \cite{hoppensteadt97,hoppensteadt99,corinto07,mirchev13,maffezzoni15,prasad22}, where several oscillators are mutually coupled through interactions and play the role of neurons. 
The basic idea in these models is to find a correspondence between targets and outputs from devices. 
For example, when one tries to associate a two-colored (black and white) pattern from memories, a correspondence between white (black) color and firing (non-firing) neuron should be given \cite{hopfield82}. 
When we perform the same phenomena using oscillator networks, a correspondence between white (black) color and in-phase (anti-phase) synchronization of the oscillators is required \cite{maffezzoni15}. 


Another effort made recently for associative memory operation is to build a virtual oscillator network \cite{tsunegi22} consisting of a spintronic oscillator, called spin-torque oscillator (STO) \cite{kiselev03}, where output from a single STO is divided into $N$ parts and treated as outputs from $N$ oscillators. 
The key point of the model \cite{tsunegi22} is that these outputs virtually interact among each other by using time-multiplexing method. 
As revealed in Ref. \cite{imai23}, the operation principle of the virtual oscillator network is similar to a feedforward neural network, rather than the conventional, instantaneously coupled oscillator networks \cite{hoppensteadt97,hoppensteadt99,corinto07,mirchev13,maffezzoni15}. 
The virtual oscillator network solved several issues in the conventional oscillator networks \cite{tsunegi22}, such as unstable operation due to inhomogeneity in the oscillators. 
However, an STO often dissipates large amount of energy because magnetization dynamics is driven by electric current. 
Therefore, it would be of great interest if we can build a similar system with different spintronics devices. 
A candidate is a ferromagnet manipulated by voltage-controlled magnetic anisotropy (VCMA) effect, where an application of voltage modulates electrons states near the ferromagnetic/nonmagnetic interface and changes magnetization direction \cite{duan08,nakamura09,tsujikawa09,maruyama09,miwa17,nozaki20}. 
Since the magnetization manipulation by the VCMA effect does not require electric current in principle, except charging and discharging the capacitor and reading, a significant reduction of the operation power is expected. 
It is, however, unclear how to develop a virtual network by employing VCMA device and perform associative memory operation. 


In this work, we propose an algorithm for an associative memory operation by manipulating VCMA effect in a single ferromagnet. 
Two stable states of the magnetization, controlled by the VCMA effect, are used as outputs of neurons in a virtual neural network. 
In this sense, the present model is similar to the original associative memory operation by neural networks \cite{hopfield82}. 
The present system, however, consists of a single device, which is the different aspect from the conventional neural networks. 
In addition, the present system is also different from the virtual oscillator network \cite{imai23}. 
This is because the algorithm developed here requires the relaxed state of the magnetization only for the computation, while the virtual oscillator network requires long-time memories of the phase during the magnetization oscillation in an STO. 
Therefore, the present algorithm is simpler than that of the virtual oscillator network \cite{tsunegi22}. 
We also demonstrate the associative memory operation of alphabet patterns. 
The applicability of the algorithm to other devices is also discussed. 



\section*{Results}

In the following, we provide a description of a virtual neural network based on VCMA device after reviewing the associative memory operation by conventional neural and oscillator networks for comparison. 

\begin{figure}
\centerline{\includegraphics[width=1.0\columnwidth]{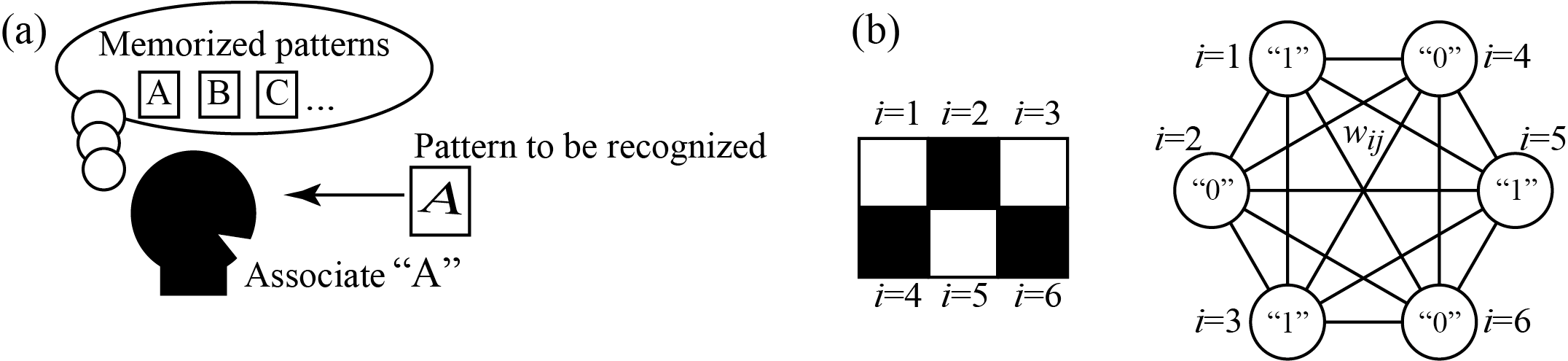}}
\caption{
         (a) Schematic illustration for definitions of association and patterns in this work. 
              A human memorizes a set of patterns, called ``memorized patterns'', such as alphabets $A$, $B$, $C$, etc.
              When a pattern, called ``pattern to be recognized'', is inputted, the human tries to find the most resembled pattern from the memorized pattern and outputs an answer. 
              This pattern recognition is an associative memory operation examined in this work. 
         (b) Schematic illustration of an example of a correspondence between a black-and-white pattern and the conventional neural network for associative memory operations. 
              The pattern is divided into $N$ pixels, and the white (black) color is related to a firing (non-firing) neuron, whose output is $1$ ($0$). 
              The neurons are fully connected to each other, and their interaction strength is proportional to weight $w_{ij}$. 
         \vspace{-3ex}}
\label{fig:fig1}
\end{figure}

The associative memory operation studied in this work belongs to a pattern recognition of a black-and-white pattern, which is schematically shown in Fig. \ref{fig:fig1}(a). 
We have a pattern, called a ``pattern to be recognized'', and aim to associate the most resembled pattern with a stored set of patterns, called ``memorized patterns''. 


\subsection*{Associative memory operation by conventional neural or oscillator networks}

Here, we briefly review the associative memory operation by neural or oscillator networks studied previously to clarify the difference between the past and present works. 
We divide patterns into $N$ pixels and also prepare neural or oscillator networks consisting of $N$ neurons, as schematically shown in Fig. \ref{fig:fig1}(b).  
An activity of the $i$th neuron is related to the color (white or black) of the pattern, as mentioned below. 

First, we need to generate a pattern to be recognized on this network, where a corresponding human activity is to see the pattern to be recognized and input it into the brain; see Fig. \ref{fig:fig1}(a). 
For this purpose, we give interactions between neurons or oscillators, which between the $i$th and $j$th neurons or oscillators ($i,j=1,2,\cdots,N$) is proportional to a weight $w_{ij}^{(1)}$. 
In this work, we use the Hebbian rule for the weight, where $w_{ij}^{(1)}$ is defined as 
\begin{equation}
  w_{ij}^{(1)}
  =
  \xi_{i}^{\rm R}
  \xi_{j}^{\rm R}.
  \label{eq:weight_1}
\end{equation}
Here, $\xi_{i}^{\rm R}=+(-)1$ when the color of the $i$th pixel in the pattern to be recognized is white (black). 
Note that the weight is unchanged even if all of the black and white colors are swapped. 
Therefore, two patterns in which all of the black and white colors are opposite should be regarded as the same pattern. 
In the neural networks, the output $x_{i}$ from the $i$th neuron is affected by the other neurons through the interaction. 
Then, we determine a threshold so that the output $x_{i}$ becomes a digital value, $x_{i}=0$ or $1$. 
In other words, we introduce a step function $\Theta[\sum_{j=1}^{N}w_{ij}^{(1)}x_{j}+b_{i}]$ [$b_{i}$ is a bias term and $\Theta(x)=0(1)$ for $x<(>)0$] as an activation function and determine the output from the $i$th neuron. 
In the oscillator network, on the other hand, the oscillators are mutually coupled through the interactions.  
As a result, the phase difference between the $i$th and $1$st oscillators, $\Delta\psi_{i}=\psi_{i}-\psi_{1}$ with the phase $\psi_{i}$ of the $i$th oscillator, often saturates to either in-phase ($\Delta\psi_{i}=0^{\circ}$) or anti-phase ($\Delta\psi_{i}=180^{\circ}$), where, for convention, we define $\Delta\psi_{1}=0^{\circ}$ to define the origin of the phases. 
Thus, in both the neural and oscillator networks, the output from the $i$th neuron becomes one of two possible values. 
When the threshold and/or the interaction strengths of these models are appropriately determined, one-to-one correspondence between the colors (black or white) of the pattern to be recognized and the outputs from the neural ($x_{i}=0$ or $1$) or oscillator ($\Delta\psi_{i}=0^{\circ}$ or $180^{\circ}$) networks is obtained. 
In this way, the pattern to be recognized is generated on the network. 

Second, we need to find the most resembled from the memorized patterns. 
For this purpose, the interaction strengths are switched to different values, which are proportional to 
\begin{equation}
  w_{ij}^{(2)}
  =
  \frac{1}{N_{\rm m}}
  \sum_{m=1}^{N_{\rm m}}
  \xi_{i}^{m}
  \xi_{j}^{m}, 
  \label{eq:weight_2}
\end{equation}
where $N_{\rm m}$ is the number of memorized patterns. 
The parameter $\xi_{i}^{m}$ is $+(-)1$ when the color of the $i$th pixel in the $m$th memorized pattern is white (black). 
Then, the outputs ($x_{i}$) from neurons in the neural network or the phase differences ($\Delta\psi_{i}$) in the oscillator network change to those of the memorized patterns most resembling the pattern to be recognized. 
As a result, the most resembled pattern appears on the network. 
It means that the association of the pattern is achieved. 


\subsection*{Manipulation method of VCMA devices}


\begin{figure}
\centerline{\includegraphics[width=1.0\columnwidth]{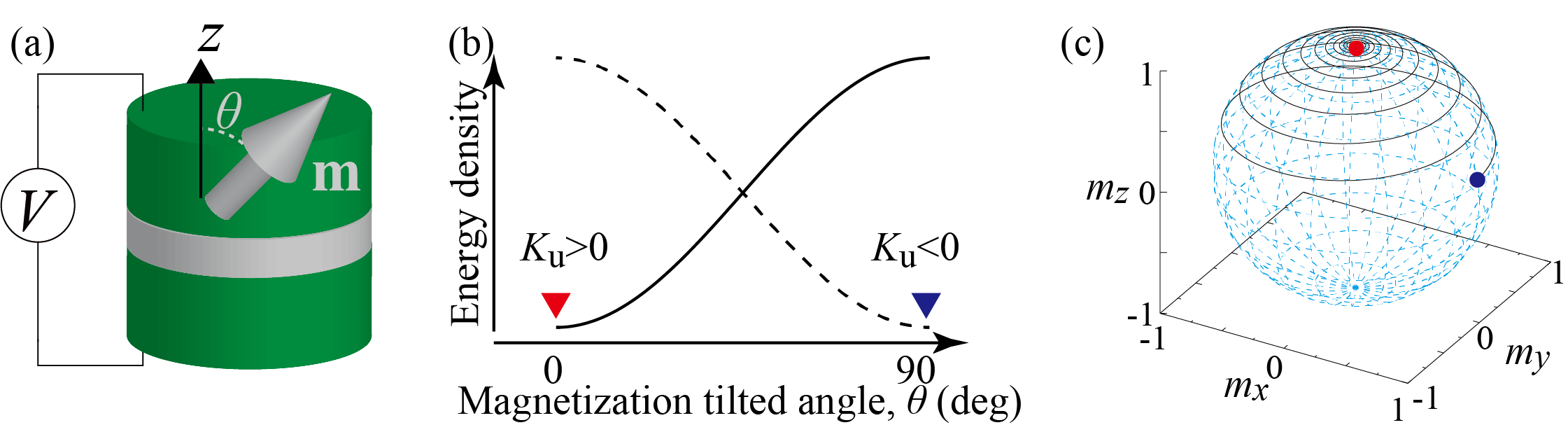}}
\caption{
         (a) Schematic illustration of a ferromagnetic/insulating barrier/ferromagnetic trilayer system with an applied voltage $V$. 
             The unit vector pointing in the direction of the magnetization in the free layer is denoted as $\mathbf{m}$, while the tilted angle of the magnetization from the perpendicular ($z$) direction is $\theta$. 
         (b) Schematic illustration of the energy density given by Eq. (\ref{eq:energy}). 
              The solid (dotted) line corresponds to the case of $K_{\rm u}>(<)0$. 
              The red and blue triangles indicate the angle $\theta$ minimizing the energy density, which is $\theta=0^{\circ}$ for $K_{\rm u}>0$ and $90^{\circ}$ for $K_{\rm u}<0$. 
              Note that the angle $\theta=180^{\circ}$ also minimizes the energy when $K_{\rm u}>0$. 
              For simplicity, however, we focus on the region of $0^{\circ} \le \theta \le 90^{\circ}$, as explained in the main text. 
         (c) An example of relaxation dynamics of the magnetization, where $K_{\rm u}>0$. 
             Since the LLG equation conserves the norm of the magnetization ($|\mathbf{m}|=1$), the change of the magnetization direction can be represented as a line on a unit sphere, as shown. 
             The initial state of the magnetization is set to be $\mathbf{m}(t=0)=(\sin 80^{\circ},0,\cos 80^{\circ})$ and is indicated by the blue circle. 
             The relaxation dynamics was evaluated by numerically solving the LLG equation, Eq. (\ref{eq:LLG}), where $\gamma=1.764\times 10^{7}$ rad/(Oe s), $\alpha=0.05$, and $H_{\rm K}=2K_{\rm u}/M=2.0$ kOe. 
             Since $K_{\rm u}$ is positive, the magnetization finally relaxes to the state $\theta=0^{\circ}$, or equivalently, $m_{z}=+1$, indicated by the red circle. 
             When $K_{\rm u}$ is negative, the magnetization will relax to the state $\theta=90^{\circ}$. 
         \vspace{-3ex}}
\label{fig:fig2}
\end{figure}


Before explaining the basic idea of the associative memory operation by a single ferromagnet, let us explain how to manipulate the output from the ferromagnetic device using the VCMA effect. 
We consider a ferromagnetic multilayer consisting of two ferromagnetic metals and one insulating barrier shown in Fig. \ref{fig:fig2}(a), where the top and bottom ferromagnets correspond to free and reference layers. 
The unit vector pointing in the direction of the magnetization is denoted as $\mathbf{m}$, where we use the macrospin assumption, which has been validated by experiments \cite{maruyama09}.  
The structure is basically the same with the STO, however, the thickness of the insulating barrier is relatively thick. 
Accordingly, while electric current flows in the STO and cause the Joule heating, the electric current, and thus the heating, are absent in the present system in principle. 
In the STO, the electric current carries spin-angular momentum and transfers it from one ferromagnet to the other, causing spin-transfer torque \cite{slonczewski96,berger96} and driving magnetization oscillation \cite{kiselev03}. 
In the present system, on the other hand, an application of electric voltage modulates magnetic anisotropy energy \cite{duan08,nakamura09,tsujikawa09,maruyama09,miwa17,nozaki20}.  
For example, the magnetic energy density of a cylinder-shaped ferromagnet consisting of the first-order magnetic anisotropy is given by 
\begin{equation}
  E
  =
  K_{\rm u}
  \sin^{2}\theta,
  \label{eq:energy}
\end{equation}
where $K_{\rm u}$ is the net magnetic anisotropy energy coefficient consisted of shape magnetic anisotropy energy, interfacial magnetic anisotropy energy \cite{yakata09,ikeda10,kubota12}, and so on. 
Importantly, the value of $K_{\rm u}$ can be manipulated by the VCMA effect and can be either positive or negative, depending on the sign and magnitude of the applied voltage \cite{maruyama09}. 
When $K_{\rm u}$ is positive (negative), the energy density $E$ is minimized when the angle $\theta(=\cos^{-1}m_{z})$ of the magnetization direction measured from the perpendicular ($z$) axis is $0^{\circ}$ and $180^{\circ}$ ($90^{\circ}$) \cite{maruyama09}, as schematically shown in Fig. \ref{fig:fig2}(b). 
When the value of $K_{\rm u}$ is changed by the VCMA effect, the magnetization changes its direction to minimize the energy density; see Methods for analytical solution of the Landau-Lifshitz-Gilbert (LLG) equation, as well as Fig. \ref{fig:fig2}(c) showing an example of the relaxation dynamics of the magnetization for $K_{\rm u}>0$. 
The magnetization state corresponding to $\theta=0^{\circ},180^{\circ}$ ($90^{\circ}$) is called a perpendicularly (in-plane) magnetized state. 
The change of the magnetization direction can be experimentally measured through magnetoresistance effect. 
Accordingly, we can generate digital output (perpendicular or in-plane) from the VCMA device by changing the sign of $K_{\rm u}$. 
If we can give a correspondence between the sign of $K_{\rm u}$ and, for example, black-and-while colors in a pattern, an associative memory operation will be possible by manipulating the VCMA device. 
For example, in the associative memory operation of black-and-white color patterns using an oscillator network \cite{maffezzoni15,imai23}, $\Delta\psi_{i}=0^{\circ}$ ($180^{\circ}$), or equivalently $\cos\Delta\psi_{i}=+1$ ($-1$), means that the color of the $i$th pixel is white (black). 
Bearing this in mind, in this work, we define the correspondence between the color and the magnetization state by parameters as 
\begin{equation}
  C_{i}
  =
  \begin{cases}
    |1 - 2\sin\theta_{1}| & (i=1) \\
    c_{1}\left( 1 - 2 \sin\theta_{i} \right) & (i \ge 2)
  \end{cases}, 
  \label{eq:C_i}
\end{equation}
where $c_{1}={\rm sign}(1-2\sin\theta_{1})$. 
Accordingly, when the magnetization reaches to an energetically stable (perpendicular or in-plane, depending on the sign of $K_{\rm u}$) state, $C_{1}$ is always $+1$, and $C_{i}$ ($i\ge 2$) is $+1$ ($-1$) when the color (black or white) of the $i$th pixel is the same with (opposite to) the $1$st pixel. 
In this work, for convention, we define that the $1$st pixel is always white, and correspondingly, define $C_{1}$ to be $+1$. 
Note that the definition of the correspondence between the angles and colors is not unique. 
For example, another possible definition is that the color of the $i$th pixel is white (black) when $\theta_{i}=0^{\circ}$ ($90^{\circ}$). 
In this case, the color of the $1$st pixel is not fixed to white. 
Such arbitrariness of the definition of the color does not affect the association because the patterns should be regarded as the same even when all of the black and white pixels are swapped. 
In addition, for convenience, we assume that $\theta$ relaxes to $0^{\circ}$ only when $K_{\rm u}>0$, in the following (see also Methods for analytical solution of the LLG equation). 

At the end of this subsection, we give two comments on the relation between the present system and previous works. 
First, comment can be made as to the possibility to extend the present system to analogue-output systems. 
While the activation function initially used in the neural network was a step function \cite{hopfield82} generating digital outputs, various kinds of activation functions, generating analogue outputs, such as hyperbolic tangent, sigmoid, and rectified linear unit (ReLU) functions, have been proposed and used in the studies of neural networks \cite{mandic01}. 
Such systems have been of great interest from both fundamental and practical viewpoints. 
In addition, in oscillator networks, interactions between oscillators sometimes result in phase differences $\Delta\psi_{i}$ which are neither in-phase nor anti-phase \cite{maffezzoni15}. 
For simplicity, however, the present work focuses on the digital outputs from ferromagnets only. 
An extension to analogue outputs will be possible if we, for example, add higher-order terms of magnetic anisotropy (see Methods for analogue output from VCMA devices). 
In fact, physical reservoir computing was studied previously by using VCMA device with second-order magnetic anisotropy energy, where analogue outputs were used for recognition task of time-dependent inputs,  \cite{taniguchi22}. 
There is also an interesting proposal for an associative memory operation of colored patterns by an array of STOs \cite{prasad22}. 
Second, we note that there is a VCMA device applicable to oscillator networks.  
Recently, a parametric oscillation of the magnetization by using microwave VCMA effect was found in Ref. \cite{yamamoto20}. 
Thus, one might consider to replace an STO in the virtual oscillator network \cite{tsunegi22,imai23} with this parametric oscillator and develop an oscillator network. 
However, the fact that an external magnetic field is necessary for this parametric oscillator \cite{yamamoto20} suggests that such method is unsuitable for practical application. 
Non-uniqueness of the oscillation phase in this parametric oscillator will also be an issue to determine outputs uniquely \cite{taniguchi23}. 
These issues should be solved if one develops an idea of implementing VCMA devices to oscillator networks. 



\subsection*{Associative memory operation by virtual network}


\begin{figure}
\centerline{\includegraphics[width=0.9\columnwidth]{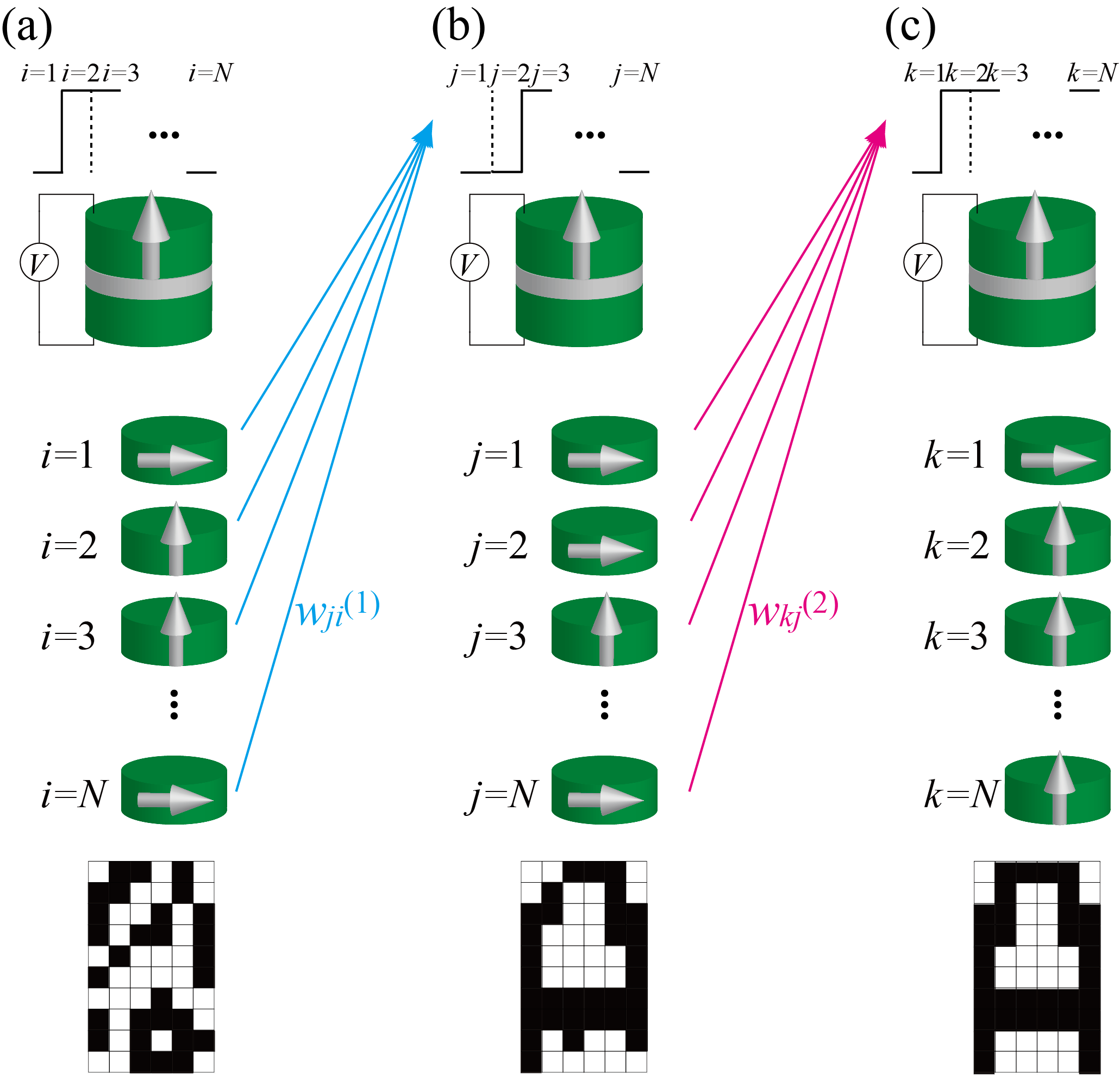}}
\caption{
         Schematic illustration of three steps for associative memory operation by a single ferromagnet manipulated by the VCMA effect. 
         The operation consists of three steps (initialization, generation of a pattern to be recognized, and association of the pattern). 
         Each step consists of applications of voltage $N$ times, and thus, is divided into $N$ parts. 
         The color of a pixel is white (black) when the saturated magnetization direction in the corresponding part is the same with (opposite to) that in the $1$st part. 
         (a) In the first step, random voltage is applied to the VCMA device $N$ times ($i=1,2,\cdots,N$), which changes the sign of $K_{\rm u}$ either to be positive or negative. 
             As a result, the magnetization during the application of the $i$th voltage relaxes to the perpendicular (in-plane) magnetized state when $K_{\rm u}$ is positive (negative), where $\theta_{i}^{(1)}=0^{\circ}$ ($90^{\circ}$). 
         (b) In the second step, voltage is applied to the VCMA device $N$ times again ($j=1,2,\cdots,N$), where $K_{\rm u}$ ends up having the value giving by Eq. (\ref{eq:K_u_1}) with $j$th voltage. 
              Here, the information of the magnetization state, $\theta_{i}^{(1)}$, with weight $w_{ji}^{(1)}$, is used to determine the $j$th voltage [or equivalently, $K_{{\rm u},j}^{(1)}$] through Eq. (\ref{eq:K_u_1}).
              Because $w_{ji}^{(1)}$ is determined by the pattern to be recognized, the magnetization direction during the application of the $j$th voltage relaxes to the magnetization state $\theta_{j}^{(2)}$ (perpendicular or in-plane) corresponding to the color of the $j$th pixel in the pattern to be recognized. 
         (c) In the third step, voltage is applied to the VCMA device $N$ times again ($k=1,2,\cdots,N$), again in which $K_{\rm u}$ results in having the value given by Eq. (\ref{eq:K_u_2}) with $k$th voltage. 
              Here, the information of the magnetization state, $\theta_{j}^{(2)}$, with weight $w_{kj}^{(2)}$, is used to determine the $k$th voltage [or equivalently, $K_{{\rm u},k}^{(2)}$] through Eq. (\ref{eq:K_u_2}).
              Because $w_{ji}^{(2)}$ is determined by the memorized patterns, the magnetization direction during the application of the $k$th voltage relaxes to the magnetization state (perpendicular or in-plane) corresponding to the color of the $k$th pixel in the most resembled pattern.  
         \vspace{-3ex}}
\label{fig:fig3}
\end{figure}


Now let us explain how to perform the associative memory operation by a single VCMA device. 
Similar to the virtual oscillator network by an STO \cite{imai23}, the associative memory operation by the virtual network consists of three steps. 
Each step consists of applying voltage $N$ times, and therefore, totally the voltages should be applied $3N$ times. 

First, we apply random voltage $N$ times, where the duration time of the voltage should be sufficiently longer than the relaxation time of the magnetization to the energetically stable state (see also Methods for the analytical solution of the LLG equation and numerical methods). 
The $i$th ($i=1,2,\cdots,N$) voltage determines the sign of $K_{\rm u}$ (positive or negative) through the VCMA effect. 
The magnetization relaxes to the perpendicular $\theta_{i}^{(1)}=0^{\circ}$ [in-plane $\theta_{i}^{(1)}=90^{\circ}$] state when $K_{\rm u}$ is positive (negative), as schematically shown in Fig. \ref{fig:fig3}(a). 
Recall that the magnetization angle $\theta_{i}^{(1)}$ ($i=1,2,\cdots,N$) can be measured through magnetoresistance effect. 
This angle $\theta_{i}^{(1)}$ is regarded as an output from the $i$th virtual neuron during the first step. 
The aim of the first step is to prepare the initial states of the $N$ virtual neurons. 
The color of the $i$th pixel in this initialized pattern is white (black) when the saturated angle of the magnetization is the same with (opposite to) that in the $1$st part, as mentioned. 
The values of the saturated angle are also stored in a memory (see also Methods for simplification of the first step). 

Second, we apply the voltage $N$ times again. 
The $j$th ($j=1,2,\cdots,N$) voltage is determined so that it makes $K_{\rm u}$ as 
\begin{equation}
  K_{{\rm u},j}^{(1)}
  =
  \frac{\tilde{K}_{\rm u}^{(1)}}{N}
  \sum_{i=1}^{N}
  w_{ji}^{(1)}
  \left[
    1
    -
    2\sin\theta_{i}^{(1)}
  \right], 
  \label{eq:K_u_1}
\end{equation}
where the weight $w_{ij}^{(1)}$ is given by Eq. (\ref{eq:weight_1}). 
Recall that $\theta_{i}^{(1)}$ obtained in the first step was either $0^{\circ}$ or $90^{\circ}$. 
Therefore, the factor $1-2\sin\theta_{i}^{(1)}$ in Eq. (\ref{eq:K_u_1}) is $+(-)1$ when $\theta_{i}^{(1)}=0^{\circ}$ ($90^{\circ}$). 
This factor, $1-2\sin\theta_{i}^{(1)}$, is similar to the color of the $i$th pixel generated in the first step, as implied from Eq. (\ref{eq:C_i}). 
The coefficient $\tilde{K}_{\rm u}^{(1)}$ determines the magnitude of the magnetic anisotropy energy during this second step. 
The numerical factor $N$ is added to the denominator to keep the value of $K_{{\rm u},j}^{(1)}$ realistic (see Methods for numerical methods).
As a result of the modulation of $K_{\rm u}$, the angle $\theta_{j}^{(2)}$ during the application of the $j$th voltage will saturate to either $0^{\circ}$ or $90^{\circ}$, depending on the sign of $K_{{\rm u},j}^{(1)}$, as schematically shown in Fig. \ref{fig:fig3}(b). 
This angle $\theta_{j}^{(2)}$ is regarded as an output from the $j$th virtual neuron during the second step. 
There will be a correspondence between the value of $\theta_{j}^{(2)}$($=0^{\circ}$ or $90^{\circ}$) and the color (white or black) of the $j$th pixel in the pattern to be recognized through Eq. (\ref{eq:C_i}). 

Third, we apply the voltage $N$ times again, which gives the value of $K_{\rm u}$ as 
\begin{equation}
  K_{{\rm u},k}^{(2)}
  =
  \frac{\tilde{K}_{\rm u}^{(2)}}{N}
  \sum_{j=1}^{N}
  w_{kj}^{(2)}
  \left[
    1
    -
    2 \sin\theta_{j}^{(2)}
  \right], 
  \label{eq:K_u_2}
\end{equation}
where $\tilde{K}_{\rm u}^{(2)}$ determines the magnitude of the magnetic anisotropy energy during this third step, while $w_{kj}^{(2)}$ is given by Eq. (\ref{eq:weight_2}). 
The magnetization during the application of the $k$th voltage will relax to either $0^{\circ}$ or $90^{\circ}$, which corresponds to the color of the $k$th pixel in the memorized pattern most resembling the pattern to be recognized, as schematically shown in Fig. \ref{fig:fig3}(c). 
Then, the association of the pattern is completed. 


For experimental researchers, let us provide a description of an experimental procedure more, although the main focus of this paper is to provide a theoretical aspect of the associative memory operation. 
The experimental equipment necessary to perform the present proposal is mainly the same with those used in typical VCMA experiments, i.e., the source meter units, probes, and so on. 
Applying voltage to adjust $K_{\rm u}$ to that determined by Eq. (\ref{eq:K_u_1}) or (\ref{eq:K_u_2}), the magnetization direction after the relaxation should be estimated through the resistance measurement.
Memory storing the information on the magnetization direction in the $i$th ($i=1,2,\cdots,N$) part during the $m$th ($m=1,2$) step and determining the voltages in the $(m+1)$th step is necessary. 
We note that a circuit generating an in-plane external magnetic field used in the switching measurement by the VCMA effect is unnecessary for the associative memory operation. 
We also note that complex measurement systems used in other spintronic neuromorphic computing, such as arbitrary-wave generator and bias-Tee used in the physical reservoir computing by STOs \cite{tsunegi18}, are unnecessary because it is unnecessary to measure, for example, an oscillating output, in contrast to coupled oscillator networks. 
In Fig. \ref{fig:fig3}, we assume step-function-like voltage inputs for simplicity, which modulate the net perpendicular magnetic anisotropy energy. 
Recall, however, that only the accurate control on the sign of $K_{\rm u}$ is necessary for the associative memory operation, and, for example, waveform of the input voltage does not affect the results of the association. 
For example, even if there is finite rising time in the voltage pulse, it does not prevent the association. 
Moreover, even if the voltages are discontinuous, i.e., there is separation time between $i$th and $(i+1)$th voltages in Fig. \ref{fig:fig3}, the present algorithm works. 
These are differences from the switching experiments utilizing the VCMA effect, where the switching probability is sensitive to the form of the voltage inputs \cite{lee17}. 
It also differs from physical reservoir computing using STOs, where it was found that the computational capability depends on the waveform of the inputs \cite{tsunegi18}. 
In these experiments, an adjustment of the waveform in nanosecond regime significantly affects the performance. 
In contrast, such a careful treatment on the inputs is unnecessary in the present proposal. 
Only the condition required for the associative memory operation is that the pulse width of the input is sufficiently longer than the relaxation time of the magnetization 
(see Methods for analytical solution of the LLG equation, which discuss the relaxation time). 


In contrast with the conventional neural networks for associative memory operation \cite{hopfield82}, the interaction between neurons in the present system is not instantaneous. 
Rather, we divide output from a single ferromagnet into $N$ parts, treat them as outputs from $N$ neurons, and give virtual interactions between them, where an output $\theta_{i}^{(m)}$ from the $i$th neuron during the $m$th step is used in the inputs $K_{{\rm u},j}^{(m+1)}$ to the $j$th neuron in the $(m+1)$th step. 
In this sense, the present system is similar to the feedforward neural network. 
However, the weights in the present system are fixed, while those in the conventional feedforward neural networks are often updated during deep-learning process. 
This is the different aspect from the feedforward neural network. 
Although we apply the voltage $N$ times during the first step mentioned above, this procedure might be simplified (see Methods for simplification of the first step).  
Therefore, the energy for the operation of the present algorithm is roughly proportional to $2N$ because a single VCMA device is driven $2N$ times. 
This energy is comparable with that required in the conventional neural and oscillator networks because $N$ neurons or oscillators in these devices are driven two times for generating the pattern to be recognized and associating the patterns. 
Note also that we use the relaxation dynamics of the magnetization to energetically stable states. 
Contrary, in the virtual oscillator network \cite{imai23}, the oscillation of the magnetization was excited in an STO, and the oscillation phase was used as outputs. 
From this aspect, the VCMA devices might poss the advantage of reducing the computational costs, compared with that using an STO, because only the information of the relaxed (final) state ($\theta=0^{\circ}$ or $90^{\circ}$) is necessary for the associative memory operation in the present scheme, while oscillating data for a long time should be stored for the virtual oscillator network \cite{tsunegi22}. 
Note also that the color of each pixel in the present scheme is uniquely determined because the angle $\theta$ is finally saturated to either $\sin\theta=0$ or $\sin\theta=1$, depending on the sign of $K_{\rm u}$. 
The color in the oscillator network might be, on the other hand, gray \cite{maffezzoni15}, i.e., neither black nor white, when the phase does not saturate to either $0$ or $180^{\circ}$. 
This again is the different aspect between the present system and the oscillator networks. 




\subsection*{Demonstration of associative memory operation}


\begin{figure}
\centerline{\includegraphics[width=0.5\columnwidth]{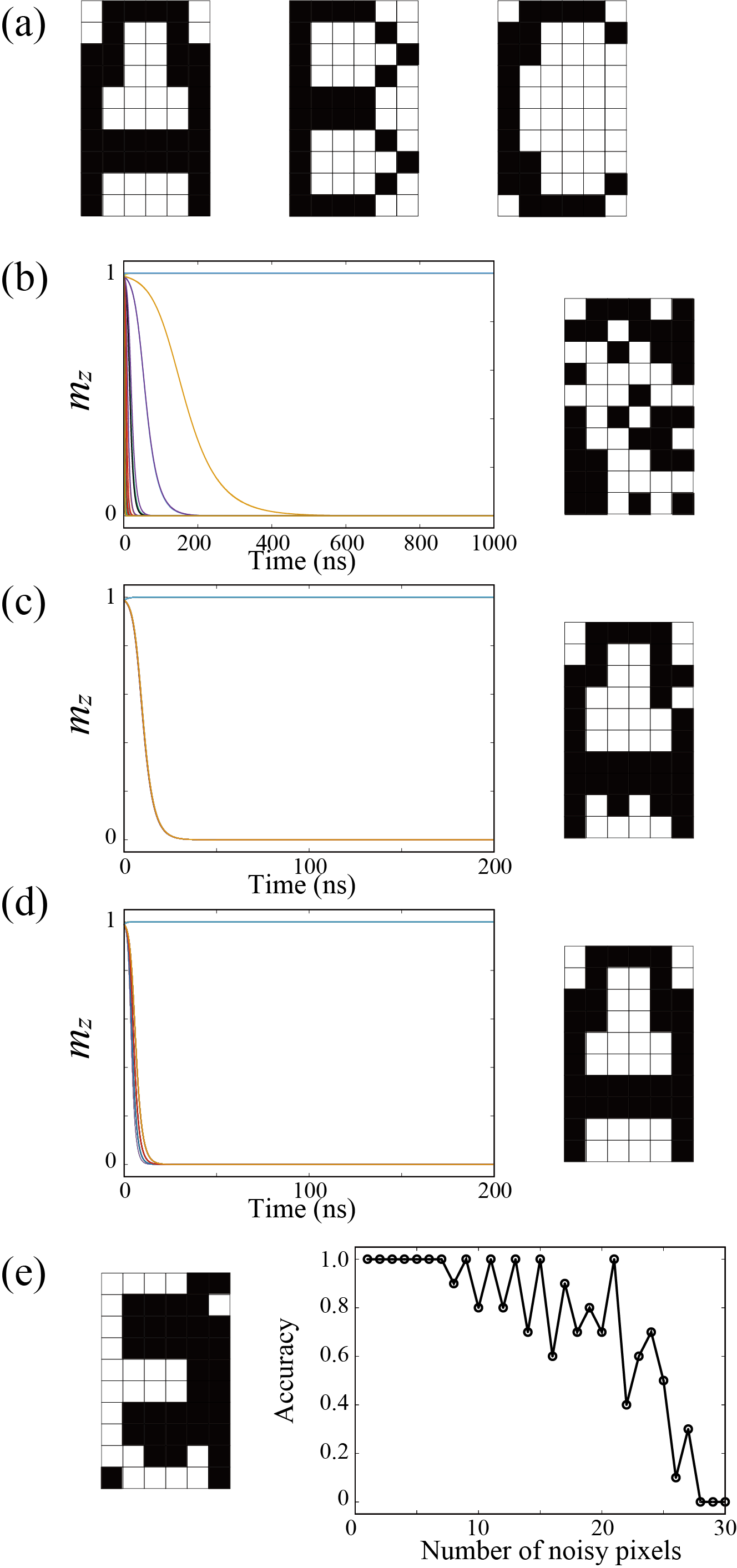}}
\caption{
         (a) Set of memorized patterns. 
         (b) Initializing the patterns, which corresponds to the first step in Fig. \ref{fig:fig3}(a).
             The left shows the relaxation dynamics of $m_{z}$ for $60$ parts and the right shows the corresponding pattern. 
         (c) Generating the pattern to be recognized, which corresponds to the second step in Fig. \ref{fig:fig3}(b).
              The generated pattern is the pattern in Fig. \ref{fig:fig5} with $4$ noisy pixels. 
         (d) Associating the most resembled pattern from the set of memorized patterns, which corresponds to the third step in Fig. \ref{fig:fig3}(c). 
         (e) An example of the pattern obtained after the third step (left) and dependence of accuracy on the number of noise (right). 
              In this case, the pattern in the left is greatly different from the pattern ``A'' in (a), and thus, the association is regarded as failure. 
              The left figure is obtained by using a figure in Fig. \ref{fig:fig5} with $26$ noisy pixels. 
         \vspace{-3ex}}
\label{fig:fig4}
\end{figure}


Here, we demonstrate the associative memory operation by a ferromagnet manipulated by the VCMA effect.
In the present work, we examine the association of an alphabet pattern from three memorized patterns, ``A'', ``B'', and ``C'', shown in Fig. \ref{fig:fig4}(a). 
Each pattern consists of $10$(rows)$\times$$6$(columns)$=60$ pixels. 
The $1$st pixel corresponds to the pixel at the upper left. 
Note that the associative memory operation for more large numbers of memorized patterns was examined in our previous work \cite{imai23}. 
As mentioned there, there had been great efforts focusing on the relationship between the maximum number of the memorized patterns and the pixel numbers, and it was found that the maximum number is approximately given by $N/(2\ln N)$ \cite{mceliece87}. 
This equation provides a rough estimation of the ability of associative memory operation by any system. 
Therefore, although we examine associations of simple patterns in this work, this equation will restrict the applicability of the present system even to real-world tasks. 
It would be of great interest to overcome this restriction not only for spintronics-based neuromorphic computing but also for general computational systems. 
We would like to keep this issue as a future work. 
 
 
As explained above, the operation uses the angle $\theta$ of the magnetization as output. 
The relaxation dynamics of the magnetization to an energetically stable state is described by the LLG equation, 
\begin{equation}
  \frac{d \mathbf{m}}{dt}
  =
  -\gamma
  \mathbf{m}
  \times
  \mathbf{H}
  +
  \alpha
  \mathbf{m}
  \times
  \frac{d\mathbf{m}}{dt}, 
  \label{eq:LLG}
\end{equation}
where $\gamma$ and $\alpha$ are the gyromagnetic ratio and the Gilbert damping constant, respectively. 
The magnetic field $\mathbf{H}$ relates to the energy density, Eq. (\ref{eq:energy}), via $\mathbf{H}=-\partial E/\partial (M\mathbf{m})$ and is given by $\mathbf{H}=H_{\rm K}m_{z}\mathbf{e}_{z}$, where 
\begin{equation}
  H_{\rm K}
  =
  \frac{2K_{\rm u}}{M}, 
  \label{eq:H_K_def}
\end{equation}
with the saturation magnetization $M$. 
Therefore, Eqs. (\ref{eq:K_u_1}) and (\ref{eq:K_u_2}) appear in the LLG equation as the magnetic anisotropy field through Eq. (\ref{eq:H_K_def}). 
The detailed relationship between $H_{\rm K}$ and Eqs. (\ref{eq:K_u_1}) and (\ref{eq:K_u_2}) in the numerical simulations, as well as the values of the parameters, are summarized in Methods for numerical methods. 
Note that Eq. (\ref{eq:LLG}) for the present system has an analytical solution for $\theta$; see Methods for analytical solution of the LLG equation. 


Recall that the associative memory operation by the virtual neural network consists of three steps, and in each step, the voltage is applied $N$ times. 
Therefore, the LLG equation should be solved $3N$ times (see also a comment in Methods for simplification of the first step). 
The point to be reminded of in the operation is that $K_{\rm u}$, or equivalently $H_{\rm K}$, varies in each part, according to Eqs. (\ref{eq:K_u_1}) and (\ref{eq:K_u_2}) (see also the Methods for analytical solution of the LLG equation and numerical methods for details). 
As a result, the solution of Eq. (\ref{eq:LLG}) has a correspondence to the pixel colors of these patterns. 


Figures \ref{fig:fig4}(b)-\ref{fig:fig4}(d) show examples of the three steps explained in the previous subsection, i.e., the initialization, the generation of the pattern to be recognized, and the association of the most resemble patterns from the set of memorized patterns, respectively. 
In each figure of Figs. \ref{fig:fig4}(b)-\ref{fig:fig4}(d), the left shows the time evolution of $m_{z}=\cos\theta$ for $N=60$ parts, while the right shows the corresponding pattern, where the color is determined by the final value of $\theta=\cos^{-1}m_{z}$. 
In the first step, the initial state of each pixel is randomly prepared, as shown in Fig. \ref{fig:fig4}(b). 
In this step, the angle $\theta_{i}^{(1)}$ is estimated from $m_{z}$ in the $i$th part and stored. 
In the second step, the pattern to be recognized is generated by using $\theta_{i}^{(1)}$ obtained in the first step and the weight $w_{ij}^{(1)}$, as shown in Fig. \ref{fig:fig4}(c). 
The reason why time evolution of $m_{z}$ are almost overlapped, although the initial condition of each part ($j=1,2,\cdots,N$) is different, will be explained in next subsection. 
Recall that the angle $\theta_{j}^{(2)}$ estimated from the $m_{z}$ in the $j$th part is stored. 
In the third step, the association of the most resembled pattern from the set of the memorized patterns is performed by using $\theta_{j}^{(2)}$ obtained in the second step and the weight $w_{ij}^{(2)}$. 
In this example, the pattern to be recognized resembles the pattern ``A'' in Fig. \ref{fig:fig4}(a), and its association is successfully achieved, as shown in Fig. \ref{fig:fig4}(d). 


Although Figs. \ref{fig:fig4}(c) and \ref{fig:fig4}(d) show a succeeded case of the association, a failure of an association possibly occurs. 
A possible origin of the failed association is that the pattern to be recognized is greatly different from any of the memorized patters (see also Methods for noisy patterns and definition of accuracy). 
Figure \ref{fig:fig4}(e) shows an example of such a failed association, where the pattern obtained after the third step is greatly different from the pattern ``A'', although the pattern to be recognized is derived from the pattern ``A'' by swapping the colors of $26$ pixels (see also Methods for noisy patterns and definition of accuracy). 
Note that whether the association is succeeded or not depends on the similarity between the pattern to be recognized and one of the memorized patterns. 
We introduce a quantity, named overlap, to quantify this similarity. 
The overlap between two patterns, $\mathscr{A}$ and $\mathscr{B}$, means the degree of having the same colors at the same pixel in these patterns. 
A quantification of the overlap can be given by 
\begin{equation}
  \mathscr{O}
  \left(
    \bm{\xi}^{\mathscr{A}}, \bm{\xi}^{\mathscr{B}}
  \right)
  \equiv 
  \frac{1}{N}
  \bigg|
    \sum_{i=1}^{N}
    \xi_{i}^{\mathscr{A}}
    \xi_{j}^{\mathscr{B}}
  \bigg|,
  \label{eq:overlap_def}
\end{equation}
where $\bm{\xi}^{\mathscr{A}}=(\xi_{1}^{\mathscr{A}},\xi_{2}^{\mathscr{A}},\cdots,\xi_{N}^{\mathscr{A}})$ is defined from the pixel color of the pattern $\mathscr{A}$ [$\xi_{i}^{\mathscr{A}}=+(-)1$ when the $i$th pixel is white (black)]. 
The overlap becomes $1$ when the two patterns are completely identical or their black and white colors are completely swapped. 
For example, the overlaps between the pattern to be recognized and the pattern ``A'', ``B'', and ``C'' shown in Figs. \ref{fig:fig4}(a) and \ref{fig:fig4}(c) are $56/60$, $32/60$, and $34/60$, respectively. 
Accordingly, the pattern ``A'' can be regarded as the most resembled pattern, and therefore, the association is regarded to be successful. 
In addition, we introduce a word ``noise''. 
In the present work, we define $w_{ij}^{(1)}$, or equivalently the pattern to be recognized by randomly swapping colors of the pattern ``A''. 
In this sense, the pattern ``A'' is regarded as an original (or target) pattern, and the aim of the associative memory operation here is to associate ``A'' with the pattern to be recognized (see also Methods for noisy patterns and definition of accuracy). 
The noise relates to the overlap as $N \left[1 - \mathscr{O}\left(\bm{\xi}^{\mathscr{R}},\bm{\xi}^{\mathscr{A}}\right) \right]$, where the symbol $\mathscr{A}$ is the index of the original (target) pattern from which the pattern to be recognized is defined (thus, $\mathscr{A}$ is ``A'' in this study). 
For example, the number of the noisy pixels in Fig. \ref{fig:fig4}(c) is $4$. 
The maximum number of noise is $N/2$ because we regard two patterns to be the same if they are obtained by swapping all the black and white colors. 


Figure \ref{fig:fig4}(e) also shows the dependence of an accuracy of the association on the number of noise (see also Methods for noisy patterns and definition of accuracy).  
Here, the accuracy is defined as follows. 
We add noisy pixels to the pattern ``A'' randomly $N_{\rm n}$ times ($N_{\rm n}=10$ in the present work).  
This pattern is used as the pattern to be recognized. 
When the pattern ``A'' is finally obtained in the third step, we regard this association accurate. 
The quantitative definition of the accuracy in the present work is as follows. 
\begin{equation}
  {\rm Accuracy}
  =
  1
  -
  \frac{1}{N_{\rm n}}
  \sum_{i=1}^{\rm N_{\rm n}}
  \frac{1}{N}
  \sum_{j=1}^{N}
  \Theta_{\epsilon}
  \left[
    |
      C_{j}^{\mathscr{A}}
      -
      C_{i,j}^{\mathscr{L}}
    |
  \right], 
  \label{eq:accuracy_def}
\end{equation}
where $\Theta_{\epsilon}(x)$ is a step function which becomes $1$  ($0$) when $x\ge (<)\epsilon$ (we use $\epsilon=10^{-3}$ in this work). 
The color of the $j$th pixel in the pattern obtained after the third step at the $i$th trial ($i=1,2,\cdots,N_{\rm n}$) is denoted as $C_{i,j}^{\mathscr{L}}$. 
Similarly, the color of the $j$th pixel in the pattern $\mathscr{A}$ is denoted as $C_{j}^{\mathscr{A}}$, where $\mathscr{A}$ represents the pattern from which the pattern to be recognized is derived. 
According to Eq. (\ref{eq:accuracy_def}), the accuracy becomes $1$ when the colors of the pattern $\mathscr{A}$ and the pattern obtained in the third step are the same with the numerical precision of $10^{-3}$. 
On the other hand, when the difference between colors in any pixels is larger than $10^{-3}$, these pixels are regarded as different, and the association of the corresponding pixel is regarded as failure. 
As the number of such differently colored pixels increases, the accuracy decreases. 
As shown in Fig. \ref{fig:fig4}(e), the accuracy falls from $1$ when the number of noise becomes larger than $\sim 10$ and becomes approximately zero when it is close to $N/2=30$. 
The role of the noisy pixel, or equivalently the overlap, on the accuracy of the associative memory operation will be discussed in the next subsection. 
Recall that two patterns are regarded as the same if all the black and white colors are swapped. 
For such a case, the minus sign in Eq. (\ref{eq:accuracy_def}) should be replaced by the plus sign. 


\subsection*{Discussion}

Now let us discuss the principle of the associative memory operation using a ferromagnet manipulated by the VCMA effect. 
As mentioned above, the associative memory operation by neural or oscillator networks consists of two processes. 
The first one is to generate the pattern to be associated on the network, and the second is to associate the pattern most resembled with it from a set of memorized patterns. 

First, we explain the operation principle to generate the pattern to be recognized on the virtual neural network consisting of a single ferromagnet. 
Recall that, before trying to generate the pattern to be recognized, we applied voltage $N$ times and obtained $\theta_{i}^{(1)}$. 
Using this $\theta_{i}^{(1)}$, Eq. (\ref{eq:K_u_1}) is defined. 
Then, the digital outputs are obtained from the ferromagnet as the angle, $\theta_{j}^{(2)}=0^{\circ}$ or $90^{\circ}$, of the magnetization, depending on the sign of $K_{{\rm u},j}^{(1)}$.  
Note that Eq. (\ref{eq:K_u_1}) can be regarded as a product of two parts as 
\begin{equation}
  K_{{\rm u},j}^{(1)}
  =
  \frac{\tilde{K}_{\rm u}^{(1)}}{N}
  \sum_{i=1}^{N}
  \xi_{i}^{\rm R}
  \left[
    1
    -
    2 \sin\theta_{i}^{(1)}
  \right]
  \times
  \xi_{j}^{\rm R}. 
  \label{eq:K_u_1_decomposed}
\end{equation}
Here, the former part, $\tilde{K}_{\rm u}^{(1)} \sum_{i=1}^{N}\xi_{i}^{\rm R} \left[ 1-2\sin\theta_{i}^{(1)} \right]$, is common throughout the second process generating the pattern to be recognized. 
It indicates that the absolute value of $K_{{\rm u},j}^{(1)}$ is the same for all of the $N$ parts, and only the sign of $K_{{\rm u},j}^{(1)}$ changes, depending on $\xi_{j}^{\rm R}$. 
Accordingly, the sign of $K_{{\rm u},j}^{(1)}$ is determined by the latter part, $\xi_{j}^{\rm R}$. 
Therefore, when the color of the $j$th pixel in the pattern to be recognized is the same with (opposite to) the $1$st pixel, the sign of $K_{{\rm u},j}^{(1)}$ is also the same with (opposite to) the $1$st pixel. 
As a result, the outputs $\theta_{j}^{(2)}$ of the second process can have a correspondence to the color in the pattern to be recognized. 
It also reveals the reason why the time evolution of $m_{z}$ in Fig. \ref{fig:fig4}(c) are almost overlapped, although we prepare $N$ different initial conditions (see Methods for numerical method). 
This is because the magnitude of $K_{{\rm u},j}^{(1)}$, or equivalently $H_{\rm K}$, determining the relaxation time (see Methods for analytical solution of the LLG equation) is common for all of the $N$ parts. 


Next, an association with the most resembled pattern in the memorized patterns is performed as follows. 
In this case, the factor $C_{j}^{(2)}$ can be replaced by $\xi_{j}^{\rm R}$. 
When the magnetization in the second step is sufficiently relaxed to the energetically stable state determined by Eq. (\ref{eq:K_u_1}) and as a result the pattern to be recognized was appropriately generated, the factor $1-2\sin\theta_{j}^{(2)}$ becomes $c_{1}^{(2)}\xi_{j}^{\rm R}$, where $c_{1}^{(2)}={\rm sign}\left[1-2\sin\theta_{1}^{(2)}\right]$. 
This replacement can be confirmed by taking into account the assumption that the color of the $1$st pixel is white ($\xi_{1}^{\rm R}=+1$), while $\theta_{1}^{(2)}$ can be either $0^{\circ}$ or $90^{\circ}$. 
Therefore, $K_{{\rm u},k}^{(2)}$ becomes 
\begin{equation}
  K_{{\rm u},k}^{(2)}
  =
  \frac{c_{1}^{(2)}\tilde{K}_{\rm u}^{(2)}}{NN_{\rm m}}
  \sum_{m=1}^{N_{\rm m}}
  \sum_{j=1}^{N}
  \xi_{k}^{m}
  \xi_{j}^{m}
  \xi_{j}^{\rm R}. 
\end{equation}
Here, we use the following approximation, 
\begin{equation}
  \sum_{j=1}^{N}
  \xi_{j}^{m}
  \xi_{j}^{\rm R}
  \simeq 
  \delta_{m,\mathscr{A}}
  \sum_{j=1}^{N}
  \xi_{j}^{m}
  \xi_{j}^{\rm R}, 
  \label{eq:sum_approx}
\end{equation}
where the symbol $\mathscr{A}$ corresponds to the one of the indexes ($m=1,2,\cdots,N_{\rm m}$) of the memorized patterns that resembles the pattern to be recognized. 
Equation (\ref{eq:sum_approx}) assumes that the most resembled pattern has a large overlap with the pattern to be recognized. 
When the overlap between the other patterns in the memorized patterns and the pattern to be recognized is small, $\xi_{j}^{m}\xi_{j}^{\rm R}$ ($j \neq \mathscr{A}$) becomes either $+1$ or $-1$, and their sum over the pixel number ($j=1,2,\cdots,N$) will be close to zero. 
This is the basis of the approximation in Eq. (\ref{eq:sum_approx}). 
When Eq. (\ref{eq:sum_approx}) holds, $K_{{\rm u},k}^{(2)}$ can be further approximated to 
\begin{equation}
  K_{{\rm u},k}^{(2)}
  \simeq 
  \frac{c_{1}^{(2)}\tilde{K}_{\rm u}^{(2)}}{NN_{\rm m}}
  \left(
    \sum_{j=1}^{N}
    \xi_{j}^{\mathscr{A}}
    \xi_{j}^{\rm R}
  \right)
  \times
  \xi_{k}^{\mathscr{A}}, 
\end{equation}
where we decomposed the right-hand side into two parts, and the former parts, $\left(\tilde{K}_{\rm u}^{(2)}/N_{\rm m}\right) \left( \sum_{j=1}^{N}\xi_{j}^{\mathscr{A}}\xi_{j}^{\rm R} \right)$, is common for all parts ($i=1,2,\cdots,N$) during the third step. 
Therefore, the color of the $k$th pixel is determined by the latter part, $\xi_{k}^{\mathscr{A}}$, which gives the color of the $k$th pixel to be the color of the most resembled pattern. 
As a result, the association among the patterns is achieved. 


The replacement of $1-2\sin\theta_{j}^{(2)}$ with $c_{1}^{(2)}\xi_{j}^{\rm R}$ mentioned above uses the assumption that the color of the $1$st pixel is white ($\xi_{1}^{\rm R}=+1$). 
One might consider a different case, where the $1$st pixel in the pattern to be recognized is black, and thus, $\xi_{1}^{\rm R}=-1$. 
In fact, when we generate the pattern to be recognized by adding noisy pixel to a pattern in the set of memorized patterns, the $1$st pixel can be coincidentally black; see Methods for noisy patterns and definition of accuracy, where $30$ examples of the pattern to be recognized are shown. 
There is also another possibility that the color of the $1$st pixel in the most resemble pattern is black. 
These cases, however, do not affect the operation principle and the evaluation of the accuracy. 
For example, when the color of the $1$st pixel in the pattern to be recognized is black, the pattern generated after the second step becomes a pattern where all of the black and white colors are swapped with respect to the pattern to be recognized. 
Since we regard such patterns identical, as mentioned, the second step should be regarded to be succeeded. 
In addition, for the third step, $1-2\sin\theta_{j}^{(2)}$ is replaced by $-c_{1}^{(2)}\xi_{j}^{\rm R}$, which results in the change of the sign of $K_{{\rm u},k}^{(2)}$. 
Then, although the saturated angles of all pixels are changed, the colors of the pattern obtained in the third step is unchanged because the colors are determined by whether the angles are the same with that of the $1$st part or not. 
Again, since we regard two patterns the same if one can be identical to the other by swapping all the black and white colors, the third step is also regarded to be succeeded. 
In addition, it is already mentioned above that the accuracy is evaluated by taking into account such a possibility, where all of the black and white colors are swapped.


According to the above discussion, whether the association becomes successful or not depends on whether there is only one pattern that has a large overlap with the pattern to be recognized. 
This is not a specific condition for the present system; rather, this has been a common and general issue for associative memory operation \cite{morita93}. 
As shown in Fig. \ref{fig:fig4}(e), when a pattern to be recognized includes large noise, the association fails. 
This is because several patterns in memorized patterns have similar overlaps with the pattern to be recognized, and the approximation used in Eq. (\ref{eq:sum_approx}) becomes no longer valid. 
An association will also be difficult when the number of memorized patterns is large and there are several patterns having large overlaps \cite{imai23}. 
It will be of interest as a future work to combine the virtual networks with deep learning and improve the success rate of the association. 


In the present work, our aim is to focus on the VCMA effect for generating digital output from a ferromagnet. 
The spin-transfer torque effect \cite{slonczewski96,berger96} will also be applicable for this scheme. 
For example, as we change the sign of $K_{\rm u}$ to change the magnetization direction, one can manipulate the magnetization direction via spin-transfer torque effect by changing the sign (direction) of electric current. 
Spin-orbit torque switching \cite{miron11,liu12} will also be applicable. 
Contrary to the virtual oscillator network utilizing an STO, such systems do not require the application of continuous electric current because once the magnetization switches its direction, the state is maintained even after the electric current is turned off. 
Therefore, it will be of great interest to develop a virtual network utilizing the VCMA or spin-transfer (or spin-orbit) torque effects from viewpoint of reducing power consumption compared with the virtual oscillator network based on the STO \cite{tsunegi22}. 
We also note that the present scheme is not limited to spintronics devices. 
This is also true for a virtual oscillator network \cite{imai23}. 
The fact that only a single device can virtually construct a network will be an interesting option for practical use because, for example, it will reduce errors due to inhomogeneities between devices and/or make the system size small. 


In conclusion, we proposed a model for the associative memory operation using a ferromagnet manipulated by the VCMA effect. 
The present model is inspired by the virtual oscillator network proposed recently \cite{tsunegi22}, which had solved several issues in the conventional oscillator networks. 
The present model has several advantages, compared with the virtual oscillator network. 
For example, using the VCMA effect will significantly reduce the power consumption due to the absence of the Joule heating, contrary to the system using STO, driven by electric current, in the virtual oscillator networks. 
The fact that only the outputs after the magnetization relaxation are used for the operation is another advantage, while the virtual oscillator network using the STO requires storing long-time data for the oscillation as outputs. 
The operation does not require external magnetic field in principle, which is preferable for practical applications. 
The applicability of the present model to the associative memory operation was confirmed by demonstrating the recognition task of alphabet patterns. 
The dependence of the accuracy in the associative memory operation on the noise in the pattern to be recognized was also evaluated. 
It was also pointed out that the present algorithm is applicable not only to VCMA devices but also to other spintronics devices, such as nonvolatile memories manipulated by spin-transfer (or spin-orbit) torque.  
Moreover, the algorithm will also be applicable to other devices. 
Therefore, this work bridges spintronics and computing science and greatly advances the applicability of spintronics technologies to neuromorphic computing. 







\section*{Methods}


\subsection*{Analytical solution of the LLG equation}

For typical VCMA devices \cite{nozaki20}, both the free and reference layers are perpendicularly magnetized. 
In this case, the experimentally measured quantity obtained through the magnetoresistance effect is the relative angle of the magnetizations in two ferromagnets (free and reference layers). 
The angle is identical to the zenith angle $\theta=\cos^{-1}m_{z}$ of the magnetization in the free layer when the magnetization in the reference layer points to the $+z$ direction, which is the case assumed in this work. 
The LLG equation for this $\theta$ is 
\begin{equation}
  \frac{d\theta}{dt}
  =
  -\frac{\alpha\gamma}{1+\alpha^{2}}
  H_{\rm K}
  \sin\theta
  \cos\theta. 
  \label{eq:LLG_theta}
\end{equation}
Integrating Eq. (\ref{eq:LLG_theta}), we find that 
\begin{equation}
  t
  =
  -\frac{1+\alpha^{2}}{\gamma\alpha H_{\rm K}}
  \ln 
  \frac{\tan\theta}{\tan\theta_{0}}, 
  \label{eq:t_sol}
\end{equation}
or equivalently,
\begin{equation}
  \theta(t)
  =
  \tan^{-1}
  \left[
    {\rm e}^{-\alpha\gamma H_{\rm K}t/(1+\alpha^{2})}
    \tan\theta_{0}
  \right],
  \label{eq:theta_sol}
\end{equation}
where $\theta_{0}$ is the initial value of $\theta$, i.e., $\theta_{0}=\theta(t=0)$. 
Note that the value of Eq. (\ref{eq:t_sol}) is not well-defined when $\theta$ and/or $\theta_{0}$ are $0^{\circ}$ or $90^{\circ}$. 
This is because the energy density,  Eq. (\ref{eq:energy}), has extreme values at these angles; when $K_{\rm u}$ is positive (negative), $\theta=0^{\circ}$ and $90^{\circ}$ attain local minimum (maximum) and maximum (minimum) of the energy density. 
It means that the gradient of the energy landscape ($\partial E/\partial \mathbf{m}$), which is proportional to the magnetic field and thus, provides torque, at these points is zero. 
Thus, the magnetization cannot move from these points. 
In reality, any perturbation, such as thermal fluctuation \cite{brown63}, moves the magnetization slightly from these points and causes the relaxation dynamics. 
We also note that $t>0$, or $\lim_{t\to\infty}\theta(t)=0^{\circ}$ ($90^{\circ}$) for a positive (negative) $H_{\rm K}$ when $\theta<\theta_{0}$ ($\theta>\theta_{0}$), which means that the magnetization relaxes to $\theta\to 0^{\circ}$ ($\theta\to 90^{\circ}$). 
The relaxation time for typical spintronic devices is on the order of $1$-$100$ ns, depending on the values of the parameters. 
Note that, if the magnetization relaxes to $\theta=0^{\circ}$ or $90^{\circ}$ immediately during an application of the $i$th voltage ($i=1,2,\cdots,N-1$), we can input next voltage. 
Having this in mind, a ferromagnet with a large damping constant $\alpha$ might be suitable for a fast computation because the relaxation time is proportional to $\alpha$. 
This point is another difference with the virtual oscillator networks \cite{tsunegi22}, where the electric current to sustain the magnetization oscillation is proportional to $\alpha$, and therefore, a small damping constant is preferable for driving an STO with low power consumption. 

Throughout the main text, we focused on two states, $\theta=0^{\circ}$ and $90^{\circ}$. 
However, we note that the state $\theta=180^{\circ}$ also minimize the energy given by Eq. (\ref{eq:energy}) when $K_{\rm u}>0$. 
For simplicity, however, we use $\theta=0^{\circ}$ only as the energetically stable state for $K_{\rm u}>0$. 
Let us give a brief comment on this point. 
In nonvolatile memory applications using the VCMA effect \cite{nozaki20}, a ferromagnet with $K_{\rm u}>0$ in the absence of voltage is used as a memory cell, and two stable states, $\theta=0^{\circ}$ and $180^{\circ}$, after turning off the voltage are used for storing information. 
In this memory devices, $K_{\rm u}$ approaches zero with the application of voltage and induces magnetization precession around an external magnetic field pointing in an in-plane direction \cite{nozaki20}. 
However, such a requirement of an external magnetic field is unsuitable for practical applications. 
In addition, the precession between two states, $\theta=0^{\circ}$ and $180^{\circ}$, occasionally becomes unstable due to high sensitivity to the voltage pulse shape and duration \cite{lee17}. 
Therefore, in this work, we consider a procedure which does not require the external magnetic field, not to mention precessional dynamics around it. 
In addition, we assume that the initial state of the magnetization locates near one ($\theta=0^{\circ}$) of two stable states; Methods for numerical methods. 
In this case, we can exclude the possibility that the magnetization arrives at another stable state, $\theta=180^{\circ}$. 
Even if the state $\theta=180^{\circ}$ is realized, we can still generate digital outputs by using $\sin\theta$ as output of the system, as in the case of Eq. (\ref{eq:C_i}) because $\sin\theta=\sin(180^{\circ}-\theta)$. 
Accordingly, we assume that the system generates digital ($\theta=0^{\circ}$ or $90^{\circ}$), not triple ($0^{\circ}$, $90^{\circ}$, and $180^{\circ}$), outputs. 
Note that, in the present scheme, we need to maintain applying voltage during the operation to keep the relaxed state and determine the output from the virtual neurons. 
This is another different aspect of the present scheme, compared with the nonvolatile memory applications, where the magnetization state after turning off the voltage is used for memory. 

At the end of this subsection, we give a more detailed definition of $K_{\rm u}$, or equivalently, $H_{\rm K}=2K_{\rm u}/M$. 
In the present work, we regard $K_{\rm u}$ as a coefficient for the net perpendicular magnetic anisotropy energy density, and as mentioned in the main text, it consists of several contributions. 
Typically, $K_{\rm u}$ is decomposed as $K_{\rm u}d=K_{\rm b}d-2\pi M^{2}(N_{z}-N_{x})d + K_{\rm i}-\eta \mathscr{E}$, where $d$ is the thickness of the ferromagnet. 
The parameter $K_{\rm b}$ represents the bulk (crystalline) contribution to the perpendicular magnetic anisotropy energy density. 
The coefficient $N_{i}$ ($i=x,y,z$) is the demagnetization coefficient, and $N_{x}=N_{y}$ for the present system because the ferromagnet is assumed as a cylinder shape.  
The term $-2\pi M^{2}(N_{z}-N_{x})$ represents the contribution from the shape magnetic anisotropy energy density. 
It becomes on the order of $1$ T in terms of the demagnetization field, $-4\pi M (N_{z}-N_{x})$. 
The parameter $K_{\rm i}$ represents the interfacial contribution to the perpendicular magnetic anisotropy energy density \cite{yakata09,ikeda10,kubota12}. 
It can also reach on the order of $1$ T. 
Accordingly, a ferromagnet can be either either perpendicularly or in-plane magnetized, as a result of the competition between the shape and interfacial magnetic anisotropy. 
The last term, $-\eta\mathscr{E}$, represents the contribution from the VCMA effect, where $\eta$ is the VCMA coefficient and $\mathscr{E}=V/d_{\rm I}$ is an electric field, where $V$ and $d_{\rm I}$ are respectively the applied voltage and the thickness of the insulating barrier separating the free and reference layers. 
The magnitude of the VCMA coefficient reaches on the order of $300$ fJ/(Vm) \cite{nozaki20APL}, which for typical VCMA devices correspond to the order of kilo Oersted. 
Summarizing them, the $H_{\rm K}$ used in the associative memory operation should be regarded as follows. 
We assume that a direct voltage is applied to canel the three contributions, $K_{\rm b}d-2\pi M^{2}(N_{z}-N_{x})d +K_{\rm i}$, and in addition to it, another voltage, which varies during the associative memory operation, is applied. 
In other words, the perpendicular magnetic anisotropy field $H_{\rm K}$, used in the numerical simulation (see also Methods for numerical methods below), should be regarded as a remaining part of the perpendicular magnetic anisotropy field after cancelling the other contributions, $(2K_{\rm b}/M)-4\pi M (N_{z}-N_{x})+[2K_{\rm i}/(Md)]$, by the VCMA effect. 


\subsection*{Analogue output from VCMA devices}

When the energy density is given as Eq. (\ref{eq:energy}), its extreme values always locate at $\theta=0^{\circ}$ and $90^{\circ}$ only. 
If we include, however, additional terms, extreme values appear at different angles. 
An example of such an additional factor is an external magnetic field $H_{\rm appl}$ whose direction is tilted from the $z$ axis with angle $\theta_{H}$. 
In this case, the energy density has an additional term, $-MH_{\rm appl}\cos(\theta-\theta_{H})$ (we assume that the magnetic field is applied in the $xz$ plane, for simplicity). 
The angle $\theta$ minimizing the total energy density locates in the region of $0^{\circ}<\theta<\theta_{H}$ ($\theta_{H}<\theta<90^{\circ}$) when $K_{\rm u}$ is positive (negative), and its value changes continuously as $H_{\rm appl}$ and/or $\theta_{H}$ changes continuously. 
Thus, analogue outputs will be generated if we use such an external magnetic field as inputs. 

For practical applications, however, it is preferable to manipulate VCMA devices without using an external magnetic field. 
Another additional term, which enables us to generate analogue outputs from VCMA devices, is higher-order magnetic anisotropy energy. 
The existence of such magnetic anisotropy has been confirmed experimentally \cite{okada18,sugihara19}. 
The energy density in the presence of the second-order magnetic anisotropy energy is given by 
\begin{equation}
  E
  =
  K_{\rm u1}
  \sin^{2}\theta
  +
  K_{\rm u2}
  \sin^{4}\theta,
  \label{eq:energy_higher_order}
\end{equation}
where we rewrite $K_{\rm u}$ in Eq. (\ref{eq:energy}) as $K_{\rm u1}$ while $K_{\rm u2}$ is the coefficient of the second-order magnetic anisotropy energy density. 
We notice that the energy density is minimized at the angle 
\begin{equation}
  \theta
  =
  \cos^{-1}
  \left(
    \pm
    \sqrt{
      1
      -
      \frac{|H_{\rm K1}|}{H_{\rm K2}}
    }
  \right),
  \label{eq:theta_stable_K2}
\end{equation}
with $H_{\rm K1}=2K_{\rm u1}/M$ and $H_{\rm K2}=4K_{\rm u2}/M$ when two conditions, $H_{\rm K1}<0$ and $|H_{\rm K1}|<H_{\rm K2}$, are simultaneously satisfied. 
While the dependence of $K_{\rm u1}$ on the voltage has been extensively studied \cite{nozaki20}, the dependence of $K_{\rm u2}$ on the voltage is still unclear \cite{okada18,sugihara19}. 
Equation (\ref{eq:theta_stable_K2}) indicates that the stable angle varies continuously as the values of $K_{\rm u1}$ and/or $K_{\rm u2}$ are varied continuously by appropriated applications of the voltage. 
Therefore, it is possible to produce analogue outputs when $K_{\rm u2}$ is finite and two conditions mentioned above are satisfied.  
In fact, the VCMA device with energy density, Eq. (\ref{eq:energy_higher_order}), was proven to be applicable to physical reservoir computing \cite{taniguchi22}, where analogue outputs were used for machine learning. 


\subsection*{Simplification of the first step}

The associative memory operation by the virtual network consists of three steps, and in each step, inputs, which was magnetic field in Refs. \cite{tsunegi22,imai23} and is voltage in this work, are injected repeatedly into the system $N$ times. 
Therefore, $3N$ inputs in total are necessary for the operation. 
However, it might be possible to simplify the first step because of the following reason. 

As discussed around Eq. (\ref{eq:K_u_1_decomposed}), the generation of the pattern to be recognized on the network is achieved because $K_{{\rm u},j}^{(1)}$ is regarded as a product of two parts and the former part on the right-hand side in Eq. (\ref{eq:K_u_1_decomposed}) is common for all ($j=1,2,\cdots,N$) parts during the second step. 
It is clear in Eq. (\ref{eq:K_u_1_decomposed}) that the angle $\theta_{i}^{(1)}$ in the first step is used in this common part. 
Therefore, even if we replace $\theta_{i}^{(1)}$ with something different, such as a constant, the conclusion that the sign of $K_{{\rm u},j}^{(1)}$ is determined by $\xi_{i}^{\rm R}$ in Eq. (\ref{eq:K_u_1_decomposed}) still holds. 
Accordingly, it will be possible to simplify, or even skip, the first step. 
This is because the first step simply corresponds to preparing the initial states of neurons in the conventional neural networks, which ideally does not affect the association. 
However, note also that we performed the LLG simulation $N$ times in the first step to emphasize the similarity between the present system and the feedforward neural networks. 
In addition, if $\sum_{i=1}^{N}\xi_{i}^{\rm R} \left[1-2\sin\theta_{i}^{(1)}\right]$ in Eq. (\ref{eq:K_u_1_decomposed}) coincidentally becomes zero due to the randomness of the voltage input during the first step, $K_{{\rm u},j}^{(1)}$ becomes zero. 
In this case, an energetically stable state is not determined uniquely, and the generation of the pattern to be recognized fails. 
Even when $K_{{\rm u},j}^{(1)}$ remains finite, if it is close to zero, a long time is necessary to saturate the magnetization to a relaxed state.  
These cases should be avoided not only for this case but also for the second and third steps. 


\subsection*{Numerical methods}

Here, we describe the numerical methods for solving Eq. (\ref{eq:LLG}). 
Although the analytical solution of Eq. (\ref{eq:LLG}) is easily obtained, as shown in Eq. (\ref{eq:theta_sol}), we numerically solved Eq. (\ref{eq:LLG}) with the fourth-order Runge-Kutta method with time increment $\Delta t=0.1$ ps. 
This is because, if one is interested in performing similar simulations with more complex systems, obtaining analytical solution cannot always be guaranteed; therefore, we developed a code for solving the LLG equation numerically. 

As mentioned in the main text, the associative memory operation consists of three steps, and in each step, the voltage should be applied $N$ times to obtain the output from $N$ virtual neurons. 
Therefore, we solve the LLG equation $3N$ times. 
In each calculation, the initial state of the magnetization was prepared by adding the effect of thermal fluctuation to the LLG equation, Eq. (\ref{eq:LLG}) and solving it with the initial condition $\mathbf{m}=+\mathbf{e}_{z}$. 
This is because, before the process of voltage application, the magnetization direction is close to an energetically stable stat, still however, slightly oscillates due to the thermal activation. 
A similar method for preparing natural initial states at finite temperature was examined in Ref. \cite{imai22}. 
The effect of thermal fluctuation can be included in the LLG equation by adding random magnetic field $\mathbf{h}$ to the magnetic field. 
The Cartesian component $h_{\ell}$ ($\ell=x,y,z$) of the random field obeys the fluctuation-dissipation theorem \cite{brown63}, 
\begin{equation}
  \langle h_{k}(t) h_{\ell}(t^{\prime}) \rangle 
  =
  \frac{2\alpha k_{\rm B}T}{\gamma M\mathscr{V}}
  \delta_{k\ell}
  \delta(t-t^{\prime}), 
  \label{eq:FDT}
\end{equation}
where $\mathscr{V}$ is the volume of the free layer, while the temperature $T$ is set to be $300$ K. 
We set $\mathscr{V}$ to be $\pi \times 50^{2} \times 1$ nm${}^{3}$, where $50$ nm and $1$ nm are typical radius and thickness of the ferromagnet used in VCMA devices \cite{nozaki20}. 
The other parameters are $\gamma=1.764\times 10^{7}$ rad/(Oe s), $\alpha=0.05$, and $M=1000$ emu/cm${}^{3}$. 
In the numerical simulation, we added 
\begin{equation}
  h_{\ell}(t)
  =
  \sqrt{
    \frac{2\alpha k_{\rm B}T}{\gamma MV \Delta t}
  }
  \xi_{\ell}(t), 
  \label{eq:random_field}
\end{equation}
to the LLG equation for preparing the initial state. 
White noise $\xi_{a}(t)$ is obtained from two uniformly distributed random numbers, $\zeta_{a}$ and $\zeta_{b}$ ($0<\zeta_{a},\zeta_{b}< 1$), by the Box-Muller transformation, $\xi_{a}=\sqrt{-2\ln\zeta_{a}}\sin(2\pi\zeta_{b})$ and $\xi_{b}=\sqrt{-2\ln\zeta_{a}}\cos(2\pi\zeta_{b})$. 
The value of $H_{\rm K}$ in the absence of the voltage application is assumed to be $2.0$ kOe. 
Accordingly, the initial states locate near the perpendicularly magnetized state, $\mathbf{m}=+\mathbf{e}_{z}$ ($\theta=0^{\circ}$). 
Note that the thermal fluctuation was added to the LLG equation only for the sake of preparing the initial states. 
During the relaxation process, the random torque was absent, for simplicity. 
We believe that it does not cause any serious change in our conclusion if $H_{\rm K}$ is sufficiently large. 
This is because the relaxed state is uniquely determined by the sign of $H_{\rm K}$ ($K_{\rm u}$), and the thermal fluctuation only gives a small oscillation around the stable state. 
Recall that when $K_{\rm u}>(<)0$, or equivalently $H_{\rm K}>(<)0$, $\theta=0^{\circ}$ ($90^{\circ}$) is stable while $\theta=90^{\circ}$ ($0^{\circ}$) is the most unstable state. 
Therefore, even in the presence of the thermal fluctuation, it is highly unlikely that the magnetization stays near $\theta=90^{\circ}$ ($0^{\circ}$) when $H_{\rm K}>(<)0$. 
This is another difference with the virtual oscillator network \cite{tsunegi22}, where a long-time memory of the magnetization oscillation should be stored for computation. 
In this case, the long-time memory includes large noise because the phase of the magnetization is always affected by the thermal fluctuation \cite{kim06}. 
From this perspective, using a relaxed state of the magnetization simply leads to a reliable operation compared with using a long-time memory of an oscillation. 



\begin{figure}
\centerline{\includegraphics[width=0.8\columnwidth]{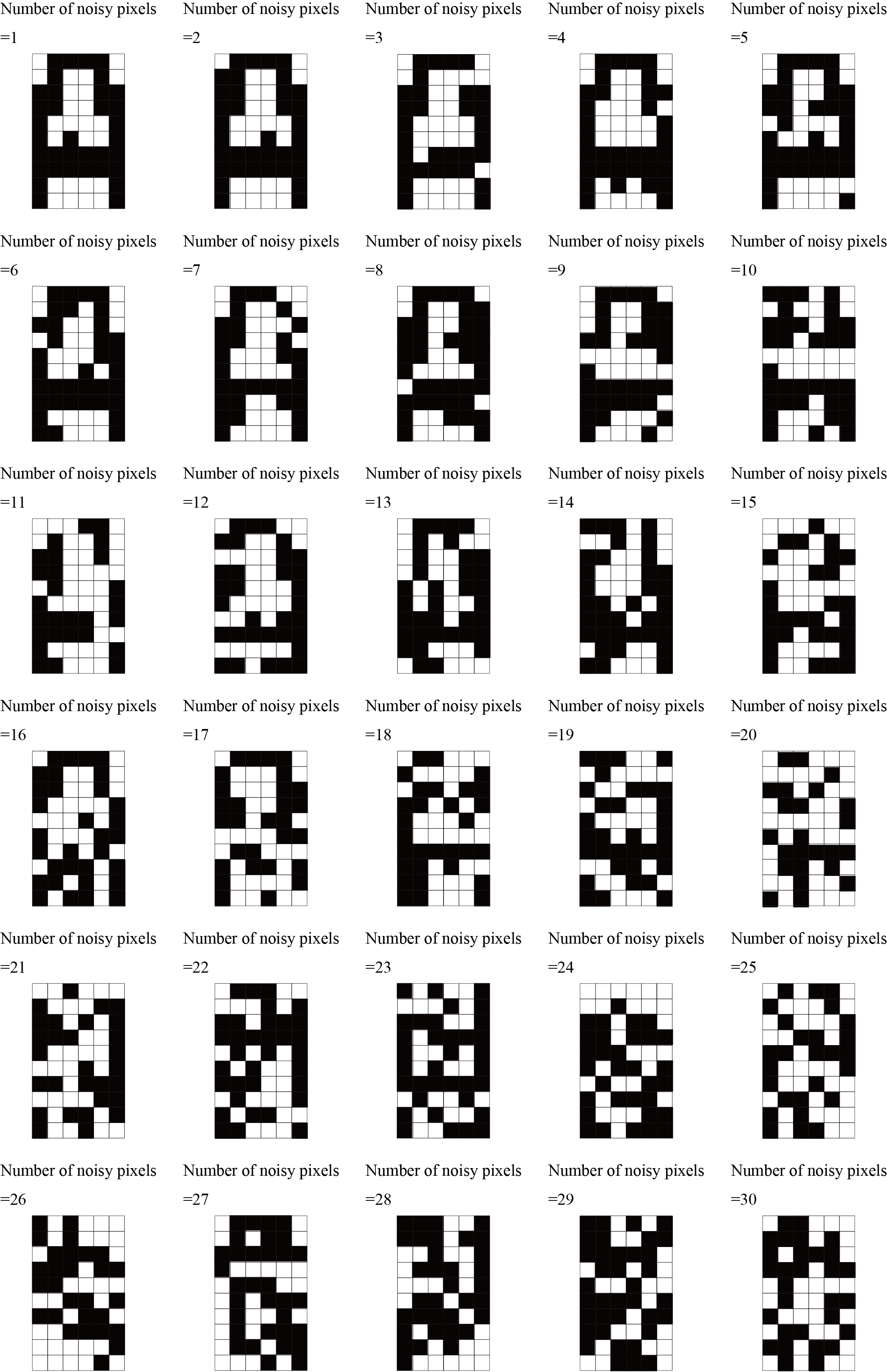}}
\caption{
             Examples of patterns to be recognized, where the number of noisy pixels increases from $1$ (upper left) to $30$ (lower right). 
         \vspace{-3ex}}
\label{fig:fig5}
\end{figure}



Next, let us explain how to give the value of $H_{\rm K}$ in Eq. (\ref{eq:LLG}) for each step. 
For the first step, $H_{\rm K}$ for each part ($i=1,2,\cdots,N=60$) is given as $H_{{\rm K},i}=H_{\rm K,0}\zeta_{i}^{\prime}$, where $H_{\rm K0}=2.0$ kOe and $\zeta_{i}^{\prime}=2\zeta_{i}-1$ with the uniform random number $0<\zeta_{i}<1$., i.e., $-1<\zeta_{i}^{\prime}<1$.  
When $H_{{\rm K},i}>(<)0$, $\theta_{i}^{(1)}$ saturates to $0^{\circ}$ ($90^{\circ}$). 
For the second and third steps, $H_{\rm K}$ for each part is $2K_{{\rm u},i}^{(1,2)}/M$, where $K_{{\rm u},i}^{(1)}$ and $K_{{\rm u},i}^{(2)}$ were introduced in Eqs. (\ref{eq:K_u_1}) and (\ref{eq:K_u_2}). 
Here, we used $2\tilde{K}_{{\rm u},i}^{(1)}/M=2\tilde{K}_{{\rm u},i}^{(2)}/M=2.0$ kOe. 
In relation to this, we give a comment on the numerical factor $N$ in the denominators of Eqs. (\ref{eq:K_u_1}) nd (\ref{eq:K_u_2}). 
Recall that $w_{ij}^{(1)}$ and $w_{ij}^{(2)}$ are $\pm 1$, and $C_{i}^{(1)}$ and $C_{j}^{(2)}$ are also $\pm 1$ after the magnetization relaxation. 
Thus, the sums of $w_{ji}^{(1)} \left[1-2\sin\theta_{i}^{(1)}\right]$ with respect to $i=1,2,\cdots,N$ in Eq. (\ref{eq:K_u_1}) and/or $w_{kj}^{(2)} \left[1-2\sin\theta_{j}^{(2)}\right]$ with respect to $j=1,2,\cdots,N$ in Eq. (\ref{eq:K_u_2}) can be on the order of $N$ at the largest. 
Accordingly, without the numerical factor $N$ in Eqs. (\ref{eq:K_u_1}) and (\ref{eq:K_u_2}), the value of $K_{\rm u}$, or equivalently $H_{\rm K}$, can become large as the pixel number $N$ increases. 
However, as mentioned above, the maximum value of the modulation of the perpendicular magnetic anisotropy in terms of magnetic field currently in consideration is on the order of kilo Oersted. 
Therefore, we added the numerical factor $N$ to the denominators of Eqs. (\ref{eq:K_u_1}) and (\ref{eq:K_u_2}) to keep the value of $H_{\rm K}$ realistic, even in the case of the large $N$. 


We should simultaneously note that the value of $H_{\rm K}$ possibly becomes significantly small. 
An example is already mentioned above for the simplification of the initial state, where the overlap between the initial state and the pattern to be recognized is small, and as a result, $K_{{\rm u},j}^{(1)}$ becomes close to zero. 
A similar thing might happen for $K_{{\rm u},k}^{(2)}$ used in the second step. 
In these cases, a long time is necessary to obtain the saturated value of the magnetization angle $\theta$. 
In the present numerical simulation, we solve the LLG equation for each part in each step for $1$ $\mu$s and estimate the angle $\theta$. 
When $H_{\rm K}$ is close to zero, $\theta$ at $t=1$ $\mu$s might differ from the saturated value determined by the sign of $H_{\rm K}$ ($K_{\rm u}$). 
This fact might affect the estimation of accuracy because the accuracy depends on the value of $\theta$ through $C_{i,j}^{\mathscr{L}}$ in Eq. (\ref{eq:accuracy_def}). 



\subsection*{Noisy patterns and definition of accuracy}

As mentioned in the main text, we add noisy pixels to the pattern ``A'' in the set of memorized patterns in Fig. \ref{fig:fig4}(a) and prepare the patterns to be recognized. 
Figure \ref{fig:fig5} show examples of these patterns to be recognized, where the number of the noisy pixels varies from $1$ to $30$. 
Recall that $30$ pixels is the maximum number of the noisy pixels, $N/2$. 
Note that the pattern with $4$ noisy pixel is the same with that used in Fig. \ref{fig:fig4}(c). 
The pattern with $26$ noisy pixels is used as the pattern to be recognized, from which the pattern in Fig. \ref{fig:fig4}(e) is obtained after performing the third step. 

We should note that the definition of the accuracy is not unique. 
An association is regarded as successful when pattern ``A'' is completely obtained in the third step. 
To generalize this definition, we introduce symbols $\mathscr{R}$, $\mathscr{A}$, and $\mathscr{B}$, where $\mathscr{R}$ represents the pattern to be recognized, while $\mathscr{A}$ and $\mathscr{B}$ represent the patterns in the set of the memorized patterns. 
We assume that pattern $\mathscr{R}$ is obtained by adding noise to the pattern $\mathscr{A}$. 
Therefore, in our definition, the association is accurate when pattern $\mathscr{A}$ is finally generated from pattern $\mathscr{R}$. 
However, when the number of noise becomes large, pattern $\mathscr{R}$ might become similar to pattern $\mathscr{B}$. 
In other words, the overlap between $\mathscr{R}$ and $\mathscr{B}$ might become larger than that between $\mathscr{R}$ and $\mathscr{A}$. 
In such a case, the pattern finally obtained after the third step will be $\mathscr{B}$. 
Even when the overlap between $\mathscr{R}$ and $\mathscr{A}$ is still larger than that between $\mathscr{R}$ and $\mathscr{B}$, the pattern obtained after the third step might be $\mathscr{B}$, depending on the number of noise. 
According to our definition of the accuracy, we regard these associations inaccurate because pattern $\mathscr{R}$ is derived from pattern $\mathscr{A}$ by adding noise. 
One might, however, regard these associations of pattern $\mathscr{B}$ accurate because nevertheless a pattern in the set of the memorized patterns is finally obtained. 
In such a case, a different definition of the accuracy is necessary. 
In the present work, we use the definition mentioned above because noise reduction (or pattern recovery) remains a challenging aspect of the associative memory operation, and our aim is to associate pattern $\mathscr{R}$ with the pattern from which it is derived, i.e., the pattern $\mathscr{A}$.








\section*{Acknowledgements}

The authors are grateful to Takayuki Nozaki for valuable discussion on the VCMA effect. 
The authors are also grateful to Takehiko Yorozu for discussion throughout the paper. 
T. T. was supported by a JSPS KAKENHI Grant, Number 20H05655. 
Y. I. was supported by a JSPS KAKENHI Grant, Number JP23KJ0331.


\section*{Author contributions statement}

T.T. designed the project, made code and performed the LLG simulations, and prepared figures.  
Y.I. made code for the generation of patterns from the data of the LLG simulations and helped the preparation of figures. 
T.T. wrote the manuscript with help of Y.I. 


\section*{Competing interests}

The authors declare no competing interests. 


\section*{Data availability}

The datasets used and/or analyses during the current study available from the corresponding author on reasonable request.


\section*{Additional information}

\textbf{Correspondence} and requests for materials should be addressed to T.T.





\end{document}